\documentclass{PoS}

\title{The $\gamma$-ray Milky Way above 10 GeV:\\
Distinguishing Sources from Diffuse Emission}

\ShortTitle{Distinguishing Sources from Diffuse Emission}

\author{\speaker{E. Owen},$^a$ C. Deil,$^{a}$ A. Donath,$^{a}$ R. Terrier$^{b}$\\
\llap{$^a$}Max-Planck-Institut f\"{u}r Kernphysik, P.O. Box 103980, D
69029 Heidelberg, Germany\\
\llap{$^b$}Astroparticule \& Cosmologie, CNRS, 75205 Paris Cedex 13, France\\
E-mail: \email{ellis.owen@mpi-hd.mpg.de}, \email{christoph.deil@mpi-hd.mpg.de}, \email{axel.donath@mpi-hd.mpg.de}, \email{terrier@apc.univ-paris7.fr}}

\abstract{One of the most prominent features of the $\gamma$-ray sky is the emission from our own Galaxy. The Galactic plane has been observed by \textit{Fermi}-LAT in GeV and H.E.S.S. in TeV light. Fermi has modeled the Galactic emission as the sum of a complex `diffuse' emission model with the predominately point source catalogs of 1FHL and 2FGL, while H.E.S.S. has primarily detected extended TeV sources. At GeV energies, Galactic diffuse emission dominates the $\gamma$-ray Milky Way but, as sources have hard spectra, it is likely their emission dominates at TeV energies. Generally the spatial shape and fraction of source emission compared to diffuse emission in the Galactic plane is not well known and is dependent on the source detection method, threshold and diffuse emission modeling methods used. \\

We present a simple image-analysis based method applied to \textit{Fermi}-LAT data from 10 GeV to 500 GeV, covering a region of +/- 5 degrees in Galactic latitude and +/- 100 degrees in Galactic longitude, to separate source and diffuse emission. This method involves elongated filter smoothing, combined with significance clipping to exclude sources. We test the method against models based on the 1FHL catalog and very simple model Galaxies to evaluate the response for an input of known fraction and shape of diffuse and source emission.}

\FullConference{Science with the New Generation of High Energy Gamma-ray experiments, 10th Workshop - Scineghe2014\\
		04-06 June 2014\\
		Lisbon - Portugal}
    
\usepackage{multirow}
\usepackage{gensymb}
\usepackage{wrapfig}
\usepackage{amsmath}
\usepackage{wasysym}
\usepackage{graphicx}
\usepackage{caption}
\usepackage{subcaption}
\usepackage{enumitem}
\DeclareGraphicsExtensions{.pdf}
\begin{document}

\section{Introduction}

In this work, we study the separation of sources from Galactic diffuse emission using image-based techniques. These are applied to data at energies between 10 and 500 GeV in order to understand appropriate methods for automatic diffuse modeling and source detection.

There is a need in the $\gamma$-ray astronomy community for a method of separating source and background `diffuse' emission in the Milky Way. In high-energy GeV and TeV data analysis, the challenge lies in finding methods that are suitable for use with the low statistics observed in this regime in order to provide reliable, stable results.

Galactic emission is thought to be the sum of a number of distinct components \cite{reimer}. However, for the purposes of this work we regard this as just two: the contribution from sources, and the background flux. In this simplified picture, sources are defined as resolvable detections which fall above a given threshold, while the remaining flux (attributed to the contributions from unresolved sources and truly diffuse emission processes such as inverse Compton, $\pi^0$ decay and bremsstrahlung) forms the background - or `diffuse' - emission.

\section{Method}

\begin{wrapfigure}{r}{0.4\textwidth}
\vspace{-10pt}
      \includegraphics[width=0.35\textwidth]{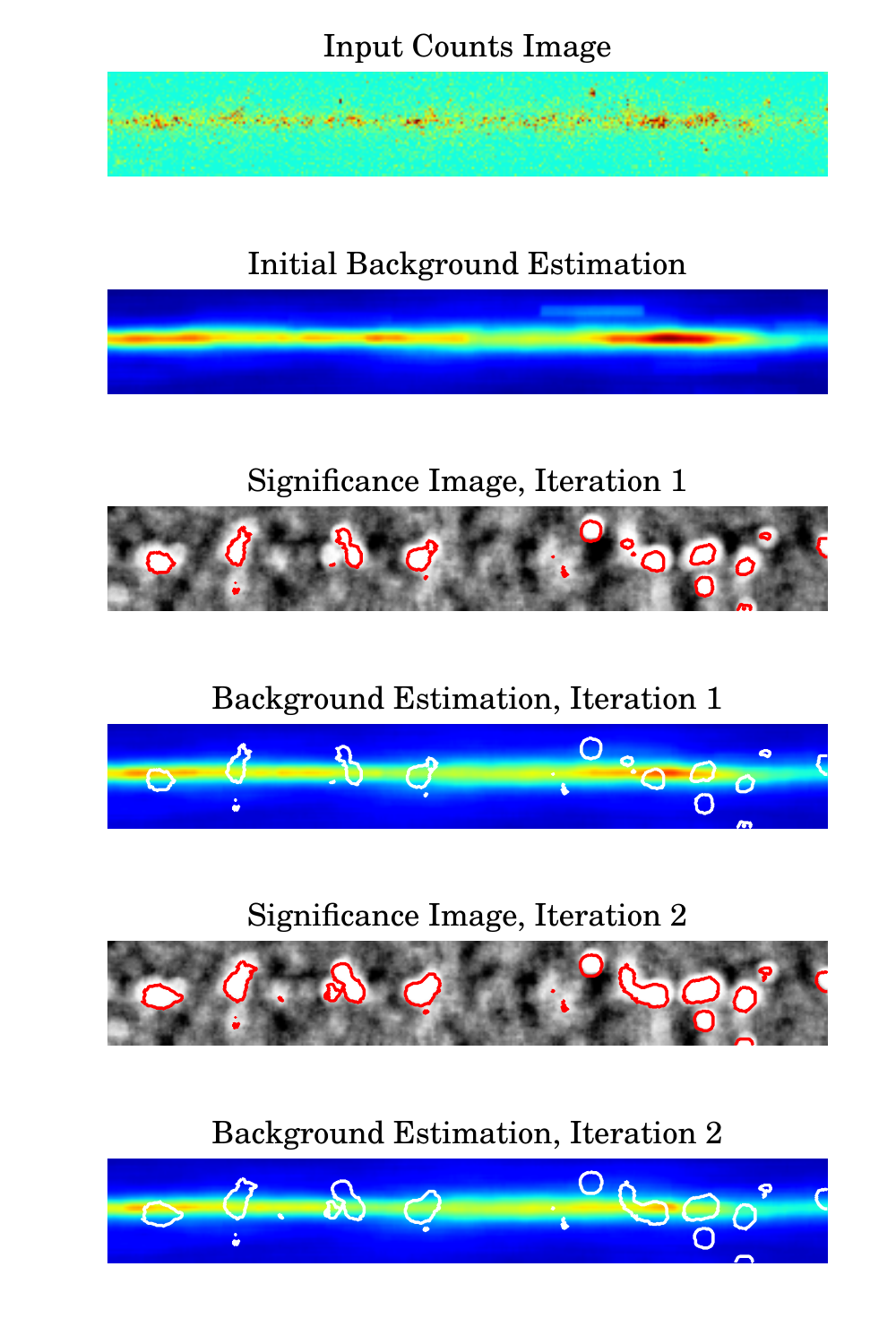}
  \caption{Iterative background method demonstration.}
\vspace{-10pt}
\end{wrapfigure}

An algorithm based on convolution with an elongated background filter and significance clipping is introduced. Its operation is summarized in the following steps.

\begin{itemize}[noitemsep,nolistsep]
\item \textbf{Background:} Convolution of a counts map with a background kernel to act as an initial diffuse background estimation.
\item \textbf{Significance:} Computation of a significance image from the counts map and background image using the Li \& Ma method \cite{LiMa}.
\item \textbf{Mask image computation:} Determination of an exclusion mask from the significance image. This removes regions above a set significance threshold which are replaced with a diffuse estimation based on the flux around the excluded source within a background convolution kernel.
\end{itemize}

The initial diffuse estimation is taken as the total flux contribution, as determined from the provided counts image. The significance and mask image computation steps repeat, with excluded regions dilated in each iteration for which they fall above the significance threshold. Once there is no change in the exclusion mask, the final diffuse estimation image is returned. This iteration process is illustrated by Figure 1, where contours indicate the shape of the exclusion mask.

Six parameters in Table 2, p. 5 (the values indicated refer to the example application to an dataset discussed in sections 3 and 4) specify the method, which can be considered a generalization of the ring background method \cite{berge} adapted to handle elongated diffuse estimation in the Galactic plane through the use of an lengthened background kernel.

\section{Dataset}

The datasets upon which the method is demonstrated comprise both of experimental observations and simulated inputs. These are intended to provide a realistic and well-understood test-base. Both are described in this section.

For all datasets, an analysis region of Galactic latitude $-5\degree < b < 5\degree$ and longitude $-100\degree < l < 100\degree$ is used to encompass the majority of the observed Galactic emission.

\subsection{Observational Data}

\textit{Fermi}-LAT data from 5 years of observation was chosen with a well defined 10-500 GeV energy cut as a showcase for the method. This choice was due to a well-understood PSF, good coverage of the Galactic plane and roughly uniform exposure of the region of interest. This region and energy band contains $6.2 \times 10^4$ events, following a power-law of index $\Gamma = -1.55$. It should be noted that applications of this method in analyses need not be limited to \textit{Fermi}-LAT data: image analysis of all high-energy Galactic data is the intended use-case.

\subsection{Simulated Data}

The simulated dataset may be split twofold: inputs for which the source population was based on observation and inputs for which a modeled source population was used. In both cases, the sources were added to a known diffuse contribution to build a complete flux image. This was taken as the most recent Fermi diffuse background model \verb|gll_iem_v05_rev1.fit|, offering a realistic distribution of background flux throughout the full study region.

Three different source populations were used. The first of these was a PSF-convolved point source image of the 1FHL catalog above 10 GeV. The resulting source-only image had a flux of $5.06\times 10^{-8} \text{ph cm}^{-2}\text{s}^{-1}$ within the study region. The remaining two source populations were entirely modeled, based on the study undertaken in the simulation work of the 1FHL Catalog paper \cite[p.59]{1fhl} which follows the essence of the study presented in \cite{Strong}. This considered a reference galaxy of a similar source population and distribution to the Milky Way, and two further simulations with different densities and proportions of bright and faint sources. Here we considered the model of similar parameters to those measured in the Milky Way and a further case where simulated parameters were chosen to produce a model of similar flux distribution to the 1FHL catalog above its detection threshold.

In line with \cite{Strong}, it was assumed that the luminosity function of the $\gamma$-ray source population between two limits $L_{\gamma, min}$ and $L_{\gamma, max}$ followed a simple power law. Further, it was taken that the spatial distribution of pulsars offered a representative sample of the full spatial distribution of a Galactic $\gamma$-ray sources \cite[p.2]{Strong}. Thus a galactocentric $(R, z)$ pulsar distribution model was employed \cite[p.7]{Lorimer} (a Gamma function) to distribute sources for the $\rho(R)$ radial density distribution function. An exponential function was used for the $\rho(z)$ distribution. Standard Monte-Carlo methods were used to sample $\rho(L_{\gamma}, R, z)$ throughout the simulated galaxy. The resulting distribution was then normalized and scaled at $R_{\astrosun}$ to the observed population density at the Sun. In line with the 1FHL study, the radial distribution was adjusted to peak at 4 kpc and the $\rho(z)$ exponential scale was set to 0.5 kpc.

A reference model (A) was defined with $\rho_{\astrosun} = 3 \text{kpc}^{-3}$, minimum $\gamma$ luminosity $L_{\gamma, min} = 10^{34} \text{ph s}^{-1}$ and maximum $\gamma$-ray luminosity $L_{\gamma, max} = 10^{37} \text{ph s}^{-1}$ with luminosity power-law index $\Gamma=-1.5$. The comparison model, B - of higher population density at the position of the Sun and lower luminosity bounds - was also considered. These are summarized in Table 1.

\begin{wrapfigure}{r}{0.4\textwidth}
\vspace{-10pt}
      \includegraphics[width=0.4\textwidth]{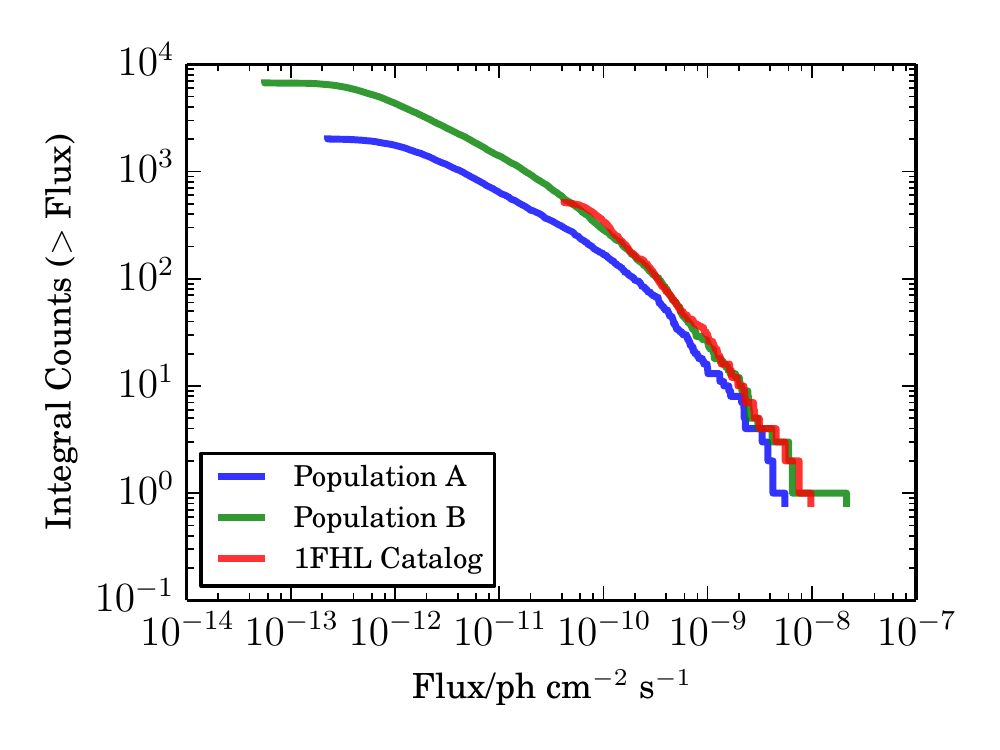}
  \caption{Simulated flux distributions.}
\vspace{-10pt}
\end{wrapfigure}

Figure 2 shows the flux distribution of each of the input galaxy populations. In the case of the 1FHL catalog, the detection threshold of order $10^{-11} \text{ph cm}^{-2}\text{s}^{-1}$ is evident. No such threshold is applied to populations A and B. Thus it can be anticipated that the separation algorithm returns better results in the case of the 1FHL source population, as there will be fewer unresolvable faint sources.

\begin{table}
\centering
\resizebox{0.8\textwidth}{!}{
\begin{tabular}{|c|c|c|c|}
\hline
\multirow{ 2}{*}{\textbf{Source Model}} & \textbf{Population Density} & \textbf{Minimum Luminosity/} & \textbf{Maximum Luminosity/}\\
 & \textbf{at the Sun/$\text{kpc}^{-3}$} & \textbf{10$^{33} \text{ph s}^{-1}$} & \textbf{10$^{36} \text{ph s}^{-1}$}\\\hline
A & 3 & 10 & 10 \\\hline
B & 10 & 4.0 & 4.0 \\\hline
\end{tabular}}
\caption{Parameters for 10 - 500 GeV Galaxy Population Simulations.}
\vspace{-20pt}
\end{table}

Before application to the source separation algorithm, the 1FHL and simulated catalogs are added to the integrated Fermi Diffuse background model \verb|gll_iem_v05_rev1.fit| between 10 GeV and 500 GeV to produce the input source \& diffuse test cases. The integral flux of this model in the Galactic plane region was $5.32 \times 10^{-7} \text{ph cm}^{-2}\text{s}^{-1}$. The fluxes of the three source populations are included in Table 3.

\section{Results}

The diffuse estimation algorithm was applied to the described inputs with the parameter values presented in Table 2. The significance threshold was set according to preliminary studies so as to avoid significance-triggering (and hence false source detection) on Poisson up-fluctuations. The correlation radius and mask dilation radius were chosen to be of the order of the \textit{Fermi}-LAT PSF to optimize for point source detection which would be observed on these scales (the LAT PSF has a 68\% containment radius of 0.150\degree and 95\% containment radius of 0.867\degree in the 10-500 GeV energy band), while an elongated background kernel of suitable size was used to ensure all exclusion regions could be covered. A pixel size of 0.1\degree offered good resolution for source detection without leading to an excessive computational load during analysis.

The results from simulated galaxy models are presented in Table 3. The performance of the algorithm can be assessed by comparing the true source flux fraction with the recovered source flux fraction. In the case of the 1FHL catalog sources more source flux is separated by the algorithm than was input, indicating some cases of source confusion. This is because the diffuse model has small-scale structure and the process of smoothing some of the peaks in the background model leads them to be falsely detected as sources in the following iteration. This is particularly evident around the galactic center. A reduction of this effect can be achieved by fine-tuning the algorithm parameters. For instance, raising the significance threshold would be expected to reduce incorrect source triggering due to diffuse background structure.

\begin{wrapfigure}{r}{0.33\textwidth}
\vspace{-10pt}
\centering
\resizebox{0.33\textwidth}{!}{
\begin{tabular}{|c|c|}
\hline
\textbf{Parameter} & \textbf{Value}\\\hline
Significance threshold & 4 $\sigma$\\\hline
Correlation radius & 0.3\degree \\\hline
Mask dilation radius & 0.3\degree \\\hline
Height of background kernel & 0.5\degree \\\hline
Width of background kernel & 10\degree \\\hline
Pixel size & 0.1\degree \\\hline
\end{tabular}}
\makeatletter
\def\@captype{table}
\makeatother
\caption{Algorithm Parameters}
\vspace{-10pt}
\end{wrapfigure}

It should be noted that there are instances in which this diffuse estimation algorithm will either fail or return incorrect results. Primarily, in cases for which the background box kernel is too small, there will be regions around sources where the exclusion mask dilates to a size larger than the box, leading to inadequate data within the kernel from which a diffuse estimate can be computed. Additionally, in cases of very low counts (\textbf{$\sim\mathcal{O}$}(10) events in the background kernel) the results will be dominated by Poisson fluctuations. In these cases it may be found that the algorithm fails entirely if significance values cannot be calculated for the full analysis region, or a null result may be returned.

For the simulated source populations, it is seen that source/diffuse separation deteriorates with as the proportion of fainter sources increases (model B has more faint sources). In these cases an effective threshold occurs where the significance of the sources against the background is not sufficient for detection. This threshold might be reduced through adjustment of the shape of the source kernel to better match the dimensions of PSF convolved faint sources, although in many cases the faintest of these sources will remain unresolvable by any method.

\begin{table}
\centering
\resizebox{0.7\textwidth}{!}{
\begin{tabular}{|c|c|c|c|c|}
\hline
\multirow{ 2}{*}{\textbf{Source Model}} & \textbf{Source Flux/} & \textbf{Total Flux/} & \textbf{True Source} & \textbf{Recovered Source} \\
 & \textbf{10$^{-8}$ ph cm$^{-2}$ s$^{-1}$} & \textbf{10$^{-8}$ ph cm$^{-2}$ s$^{-1}$} & \textbf{Flux Fraction/\%} & \textbf{Flux Fraction/\%}\\\hline
1FHL Catalog & 5.1 & 58 & 9 & 16 \\\hline
A & 8.5 & 62 & 14 & 13 \\\hline
B & 12.9 & 66 & 20 & 10 \\\hline
\end{tabular}}
\makeatletter
\def\@captype{table}
\makeatother
\caption{Galactic Plane recovered diffuse fluxes}
\vspace{-15pt}
\end{table}

Application to true \textit{Fermi}-LAT data to develop a diffuse model (Figure 3) produced a diffuse estimation with a flux of $5.17 \times 10^{-7}$ ph cm$^{-2}$ s$^{-1}$ in the Galactic region. Figure 4 presents spatial profiles of the resulting diffuse model. It can be seen that the diffuse estimation here largely follows the total flux distribution of the Milky Way with some sources excluded, notably Vela at -2.78\degree latitude and -96.4\degree longitude. Shortcomings are evident where diffuse estimation exceeds the total flux. This arises due to an inappropriately large size of the background convolution kernel in such regions, combined with source contamination in the diffuse estimation.

\begin{figure}
  \begin{center}
      \includegraphics[width=0.9\textwidth]{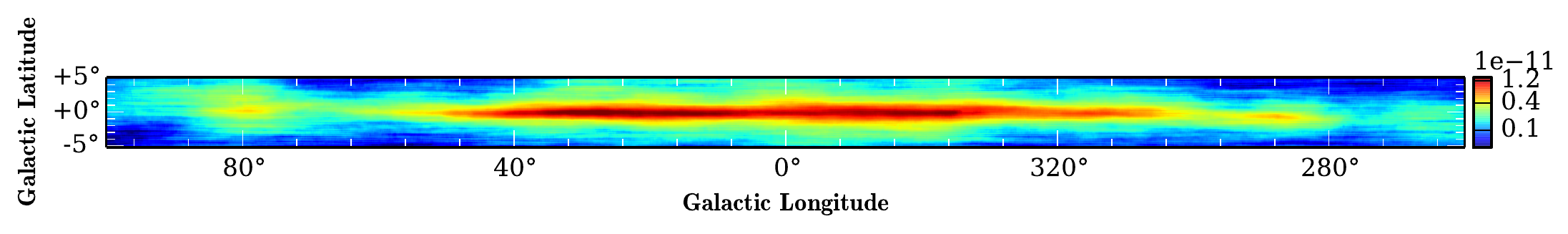}
  \caption{Background estimation for \textit{Fermi}-LAT Data.}
  \end{center}
  \vspace{-30pt}
\end{figure}

\begin{figure}
\centering
\begin{subfigure}{0.33\textwidth}
      \includegraphics[width=\textwidth]{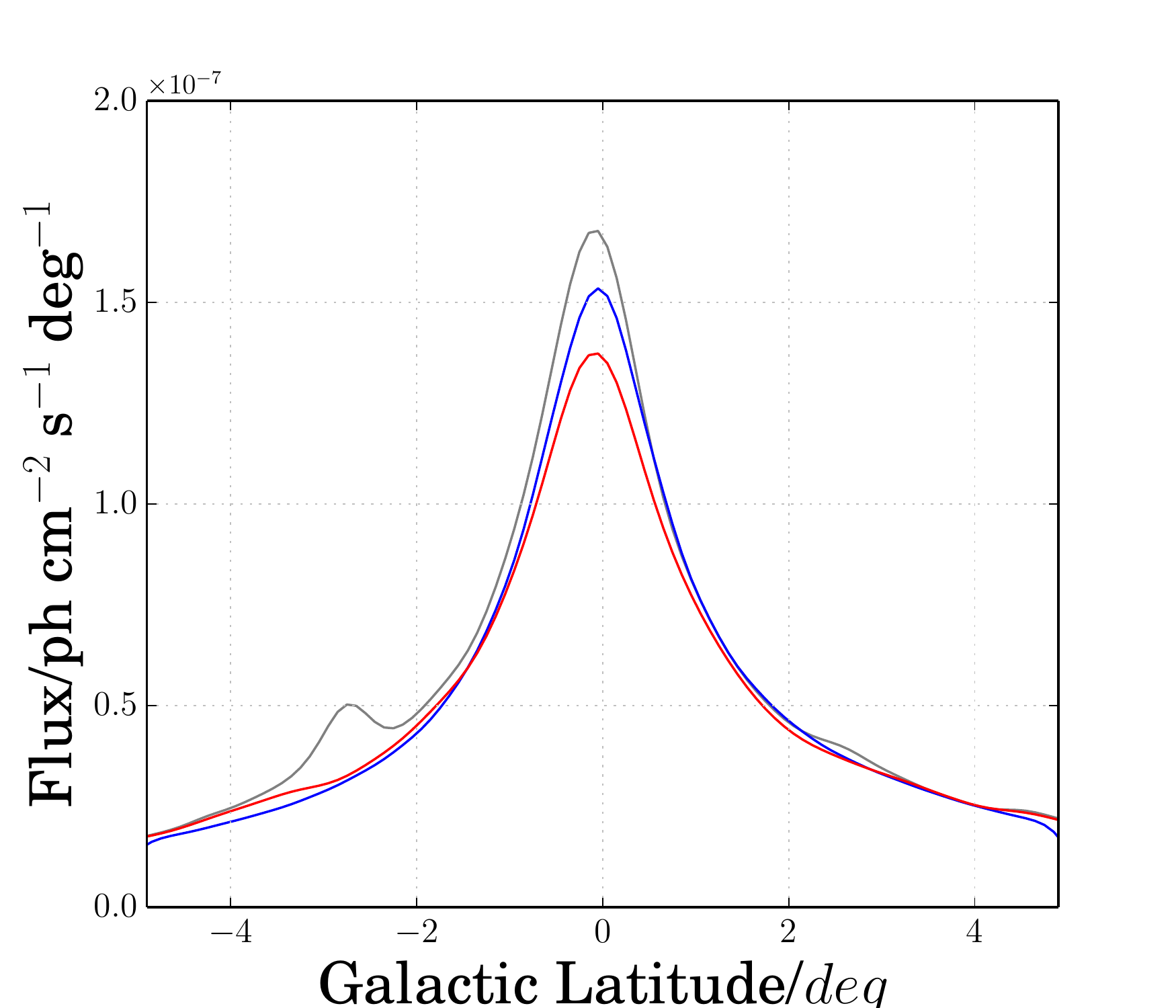}
\end{subfigure}
\begin{subfigure}{0.66\textwidth}
        \includegraphics[width=\textwidth]{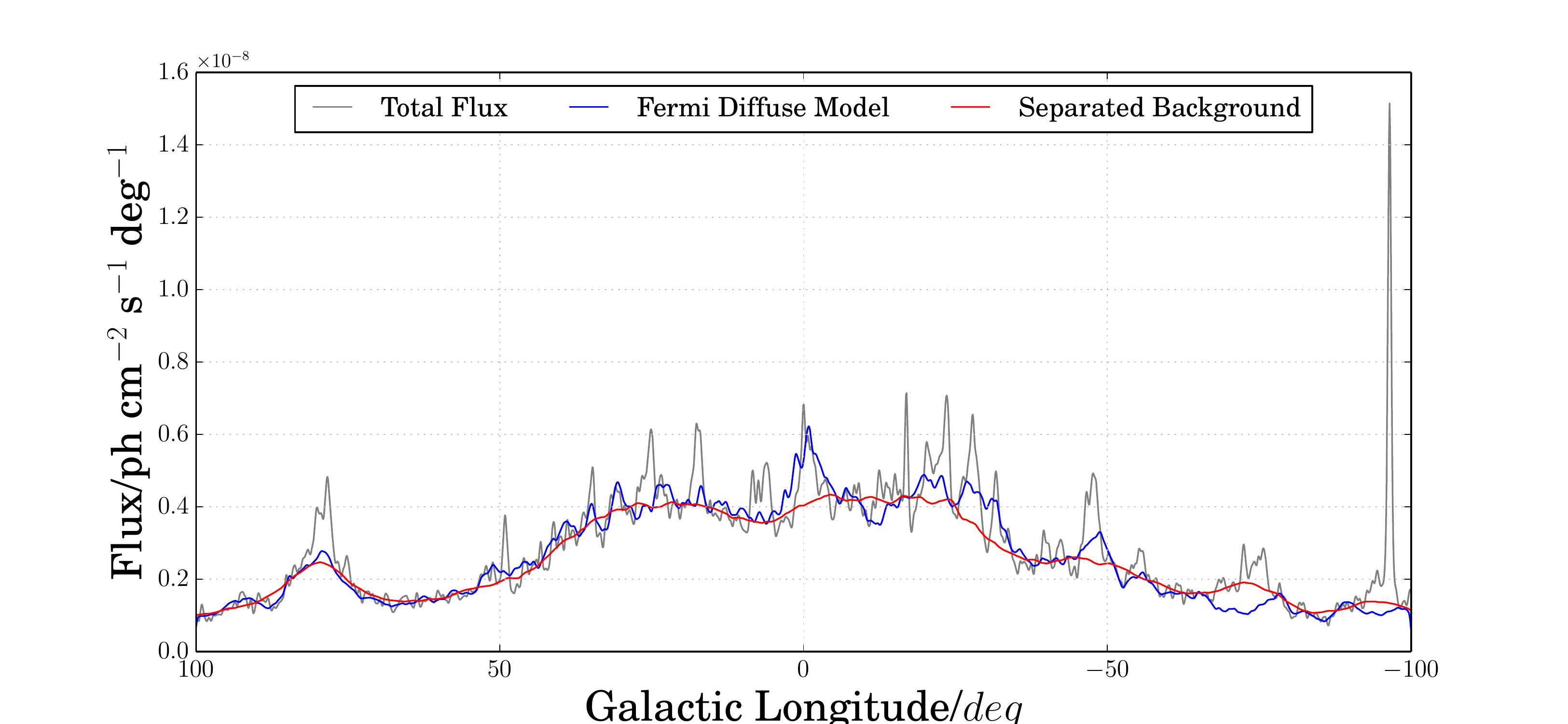}
\end{subfigure}
\caption{\textit{Fermi}-LAT Profiles for total Galactic flux and estimated diffuse flux}
  \vspace{-15pt}
\end{figure}

\section{Conclusions}

This work outlined a new method of modeling the Galactic diffuse emission and presented a diffuse emission model based on Fermi-LAT data between 10 and 500 GeV. This model is built only from the observed data, with the assumption that it is smooth on longitude and latitude scale given by the background kernel. This is fundamentally different from the diffuse emission model usually used for Fermi-LAT data analysis \cite{ackermann}, which is based on a much more complex physical model of cosmic rays and emission processes in the Milky Way \cite{reimer}.

Overestimation of the diffuse flux with this method was evident in some regions of resulting background estimations, and steps are suggested for fine-tuning. Combined with algorithm exten- sions to allow for the iterative use of multiple background kernels, improvements in the resulting diffuse estimation may be achieved. An extension of this could be to generalize the algorithm for multi-scale detection by computing significance images with multiple correlation radii such that source exclusion would occur for a wider range of morphologies, not just point like sources and those with a small angular size. The methods and algorithm introduced in this study, including the Galactic population simulation methods, are provided in the affiliated Astropy open-source package, Gammapy \cite{Deil}.

\end{document}